# Measuring the Transmission Matrix of a scattering medium using Epi-Fluorescence light

EVOLENE PREMILLIEU,[1] RAFAEL PIESTUN[1] *

[1] *Department of Electrical, Computer, and Energy Engineering, University of Colorado at Boulder, USA.*

*Corresponding author: rafael.piestun@colorado.edu*

**There are several methods to focus light behind a scattering medium, but very few use fluorescence light as feedback or can be used without access to the distal side of the scatterer. Among all the wave-front shaping techniques, retrieving the transmission matrix of a scattering material is the only one that allows for focusing on multiple spots after a single set of measurements. Here we propose a method to retrieve the transmission matrix of a scatterer using fluorescence light as feedback without access to the distal side. The advantage of this method is that it allows focusing on the whole field of view after a single matrix measurement without affecting the sample, making it suitable for reflection epi-illumination geometry microscopy.**

**Introduction.** Fluorescence microscopy is widely used for a diversity of samples due to its high contrast, capability of targeting specific structures, and minimal toxicity, which enable dynamic, structural, and molecular interaction studies. However, when it comes to thick or highly scattering samples, the fluorescent signal is degraded. Confocal and multiphoton microscopy give good resolution but depend on ballistic light to focus through the scattering layers [1,2]. When only a speckle pattern reaches the target, wavefront shaping offers a promising opportunity. Several methods have been developed to create a focus behind a turbid layer using the output speckle field as feedback [3] in an iterative process. Alternatively, the transmission matrix (TM) approach characterizes the medium with a set of input-output measurements that can be used to generate spots over a wide field of view [4]. In particular, scattered light fluorescence microscopy has shown sub-wavelength imaging resolution [5]. The method uses the transmitted speckle field as feedback to create a focus and by scanning it, within the so called memory effect range [6], images small structures behind highly scattering materials [5,7]. By adding a curvature on the optimized phase mask, scanning within the memory effect range in three dimensions has been demonstrated as well [8]. Both transmission and reflection geometry have been studied [3,5,7,8], paving the way to imaging behind or inside turbid media. Another interesting approach combines the advantages of structured illumination microscopy and wavefront shaping to generate a light-sheet behind a turbid medium, enhancing the signal of fluorescent beads by a factor of ~8 [9]. Most demonstrations of these wavefront shaping methods have provided very good resolution behind a scattering medium, but they only allow for imaging of a very small field of view (FOV). Structured illumination behind a scattering layer addresses this issue by enabling the creation of multiple foci that can each be scanned within the memory effect range [10].

The standard method for retrieving the optical transmission matrix of a scattering medium involves projecting a set of input functions with a spatial light modulator (SLM) and measuring the interference of the output speckle field with a phase shifted reference [4]. There are two options to implement the reference field: through an external reference arm or through an internal reference using part of the SLM active area. While the former takes advantage of all the degrees of freedom of the SLM and generates a uniform reference, the latter is more stable and provides only a speckled reference [11]. The transmission matrix can also be retrieved without a reference arm, using phase retrieval algorithms based on several calibrations [12]. There have been more variations of the TM measurement, for which, in all cases, the need for coherent light limits its fluorescence applications to transmission geometry set-ups.

Unfortunately, most fluorescence microscopy methods operate in reflection geometry (epi-fluorescence). Hence, they require performing shaping in reflection, without direct access behind the scattering sample. While it has been demonstrated that fluorescence feedback can be used for focusing on a single fluorescent bead, the limitation to the small field of view remains a problem [8]. It should be noted that reflection matrix measurements using coherent (non-fluorescent) backscattered light [13,14,15] would not necessarily perform in fluorescence.

Here we propose a method to retrieve the transmission matrix of a scattering medium using fluorescent light as feedback. As a result, it enables focusing on multiple beads with a single TM measurement. Our method is inspired by work done in the photoacoustic TM measurement [16], except that here we use fluorescence intensities instead of photoacoustic signal amplitudes. Importantly, the process does not require access to the back of the scattering medium, which offers new possibilities for reflection geometry applications. The main requirement for the technique is that the beads generate enough signal to noise ratio (SNR) to reflect the change in fluorescence intensity before the scattering surface and that there is weak mixing among the scattered fields originating from the different beads. We show that we can increase the signal of several beads by up to a factor of 15 with a single TM measurement. To compare and validate our results, we also use a genetic algorithm (GA) to focus on a single bead. Genetic algorithms have been shown to be robust in noisy environments [17]. We obtained similar enhancement results for the same sample, which confirms the potential interest of using the TM approach.

**Theory and method.** The transmission matrix is the relation between the output fields and the input optical modes, mathematically expressed as $E_m^{out} = \sum_{n=1}^{N_{SLM}} T_{mn} E_n^{in}$, where $N_{SLM}$ is the total number of active modes, typically the number of pixels as input on a spatial light modulator (SLM), while $m$ is the camera pixel index. Here, to measure the transmission matrix, we use the Hadamard basis with the number of patterns equal to the number of SLM pixels, so the index $n$ corresponds to the Hadamard pattern index. Our method is based on the conventional phase-shifting interferometric method used to retrieve the TM [4]. We use an internal reference, consisting of a frame around the Hadamard patterns projected on the SLM. The phase of the frame stays fixed for these experiments and we shift the phase of the Hadamard patterns between 0 and $2\pi$ in 8 steps. The process for acquiring the TM is depicted in Fig.1. As the phase is varied on the SLM, the detected fluorescence intensity is proportional to the linear sum of all the input modes $m'$ reaching the fluorescent bead [16]:

$$I_{Fluo} \propto \sum_{m'} \left| E_{m'}^{ref} + \sum_{n=1}^{N} T_{m'n} E_n^{in} \right|^2 \quad (1)$$

During the TM measurement we have $E_n^{in} \propto e^{i\varphi_n^{SLM}}$ as the input modes are phase shifted between 0 and $2\pi$ with $\varphi_n^{SLM}$ the phase on the SLM. The detected fluorescence intensity for mode $n$ is therefore proportional to:

$$I_{Fluo,n} \propto \alpha_{mn} + \beta_{mn} cos(\theta_{mn} + \varphi_n^{SLM}) \quad (2)$$

where $\alpha_{mn} = \sum_{m'}(|E_{m'}^{ref}|^2 + |\sum_{n=1}^{N} T_{m'n}|^2)$,
$\beta_{mn} cos(\theta_{mn}) = \sum_{m'}(2|E_{m'}^{ref} \sum_{n=1}^{N} T_{m'n}|)$,
$\beta_{mn} sin(\theta_{mn}) = \sum_{m'}(-2i|E_{m'}^{ref} \sum_{n=1}^{N} T_{m'n}|)$

The detected fluorescence intensity follows a cosine modulation as the phase on the SLM is shifted. The TM for each camera pixel of the field of view is simply retrieved after a discrete Fourier transform as follows:

$$T_{mn}^{Fluo} = \beta_{mn} e^{i\theta_{mn}} \quad (3)$$

With $\beta_{mn} = |T_{mn}^{Fluo}|$ and $\theta_{mn} = arg\{T_{mn}^{Fluo}\}$ as:

$$\theta_{mn} = -arg\left\{\sum_{k=1}^{Phase-steps} I_n^k \times e^{i\frac{2\pi k}{Phase-steps}}\right\} \quad (4)$$

$$\beta_{mn} = \left|\sum_{k=1}^{Phase-steps} I_n^k \times e^{i\frac{2\pi k}{Phase-steps}}\right| \quad (5)$$

After retrieving the full TM for every pixel of the field of view, focusing is performed by projecting the conjugate transpose of the TM on the SLM:

$$\varphi_n^{SLM} = -\theta_{mn} = -arg\{T_{mn}^{Fluo}\} \quad (6)$$

During the focusing part, a focus, centered on the pixel whose corresponding row of the TM is projected, is scanned across the region of interest. The detected intensity of the fluorescent beads is enhanced as the scattering is compensated. We define the back fluorescence enhancement as the ratio of the intensity after optimization over the initial intensity with a flat phase on the SLM in a given region of interest. This should be distinguished from the focused spot enhancement on the distal side of the scatterer, defined as the ratio of the peak focus laser intensity and the average intensity before optimization [17].

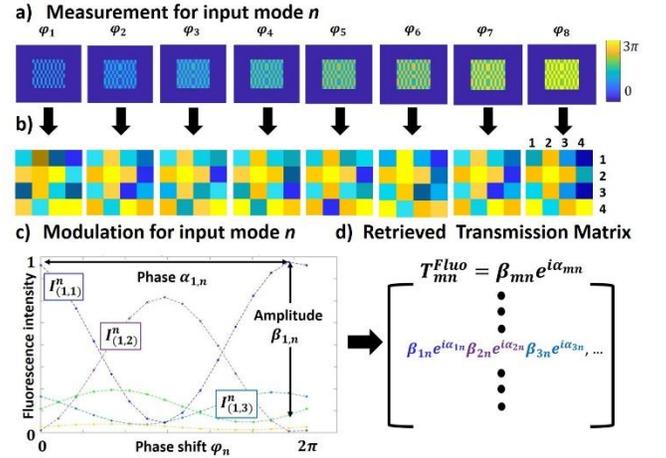

Fig. 1. **Transmission matrix measurement with epi-fluorescence light**. (a) Projection of Hadamard pattern for input mode $n$ with modulated phase between 0 and $2\pi$ in 8 steps, (b) recording of fluorescence intensity for each modulated pattern, (c) the fluorescence intensity as a function of the phase shift for input mode $n$ follows a cosine (simulation data shown), (d) retrieving the transmission matrix for input mode $n$.

The experiments were performed with the set-up represented in figure 2. A polarized laser beam (Spectra-Physics Mosaic, pulsed 532 nm, 10ns, 20kHz repetition rate) is incident on a SLM (Meadowlark, 512x512 pixels), then resized and goes through a dichroic mirror (Semrock FF541_SD:01). The SLM plane is imaged onto the sample surface with a microscope objective (Leitz, 32x, 0.6NA). For this proof of principle experiment, we used a frosted glass slide of thickness 1mm as the scatterer. Nile red beads (Spherotech 2 µm) are deposited on the other side of the ground glass surface. They are imaged with a camera (Andor, EMCCD iXon 3) and their scattered intensity is used as feedback signal for the TM measurement. The imaging lens was chosen to respect Nyquist sampling when the beads are imaged without a scattering surface, in this case we imaged one bead on 3.75 pixels. When imaged through the scatterer, the beads appear bigger due to the diffusive blurring. The exposure time was typically 0.1s.

**Experimental results.** In the experiments, 64 or 256 active SLM pixels were enough to measure the TM and obtain a back fluorescence enhancement of up to 15.8 with an average of 8.5 over 10 different measurements. The TM was measured for a region of interest containing a few beads with sizes equal or less than 50x50 pixels (limited by the hardware capability). The pixels in between the beads do not result in an accurate TM measurement as they are dimmer. However, the focus created is centered on one pixel but covers an area of one speckle (about the size of the beads in our case), allowing for scanning in between the beads.

To minimize computational cost, we chose a small FOV and focus on one bead or two at a time, with the same TM measurement. Figure 3 shows a result for two beads where the initial fluorescence signal was increased by a factor of 8.7 after focusing. For this run we used 64 active SLM pixels. The other experiments were run with 256 active SLM pixels. The TM measurement takes tens of minutes and depends on the SLM refresh rate and exposure time on the camera.

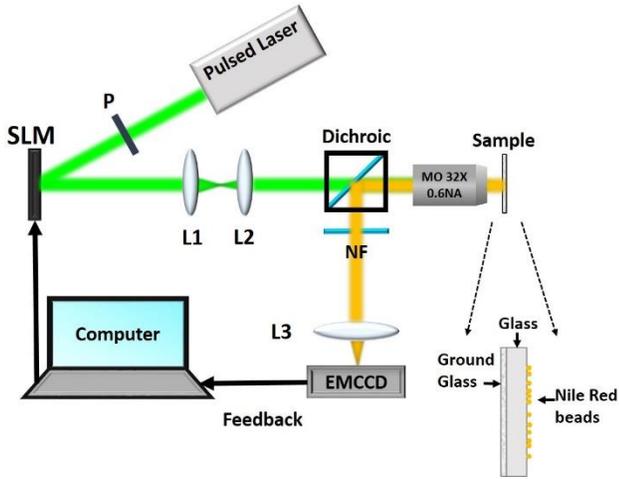

Fig. 2. **Schematic of experimental set-up**. A polarizer (P) matches the laser beam polarization to the SLM. The beam is then resized through lenses L1 and L2 (F1= 100mm, F2 = 150mm) to fill the back aperture of a microscope objective (Leitz, 32x, 0.6NA) and illuminates the scattering surface of the sample (frosted glass microscope slide with 2μm Nile Red beads (Spherotech) on the other side). A dichroic mirror separates the incident illumination from the collected fluorescence light and a notch filter (NF) blocks the remaining 532nm light from reaching the detection camera (Andor EMCCD) (imaging lens L3 with F3=150mm).

The first pixel of the region of interest we focus on is not enhanced due to the initialization of the camera in the acquisition code (see Fig.3. (b)). The enhancement map is built pixel by pixel after saving the intensities obtained when projecting the TM conjugate transpose for each pixel of the region of interest. Figure 3 illustrates the focusing process. The focus created is centered on the pixel marked with a black +, illuminating one bead while the signal of the other bead is not enhanced, which shows the localization capability of the method.

We can monitor the speckle field with a second imaging system facing the laser beam. For this series of experiments, the speckle size is typically about 2 μm at full width half maximum (FWHM), measured by speckle auto-correlation.

We use a weakly scattering medium for these experiments so that we obtain a good SNR and we do observe some ballistic photons before optimization. The size of the focus formed is about the size of one speckle, whereas the ballistic photons are spread out.

To confirm the performance of the TM approach in terms of fluorescence signal enhancement, we compare our results with the results obtained with a genetic algorithm (GA). GA are known to be robust, even in noisy conditions [17]. However, optimizing the fluorescence signal with a GA results in a single focus spot, on a single bead in our experiment. GA optimizations were performed in same conditions as TM measurements and focusing. We used 256 active SLM pixels and a population of 30 random phase masks. The acquisition time on the EMCCD is the same as for the experiments with the TM approach for a given sample. Figure 4 shows results obtained with the TM approach and the GA on the same beads. In the TM approach, a single TM measurement allows for focusing on any of the beads whereas focusing with the GA requires several successive optimizations. The final image obtained after optimization with the GA is a single acquisition while projecting the optimized phase mask on the SLM. The optimization time with the GA is comparable to the TM measurement time for equal exposure time on the camera. The enhancement of the beads' signal was comparable in both cases. For 10 optimizations with the GA, the average enhancement was 8.2. We can conclude that the TM approach is comparable to iterative optimization algorithms and offers the advantage to focus on several spots after a single measurement.

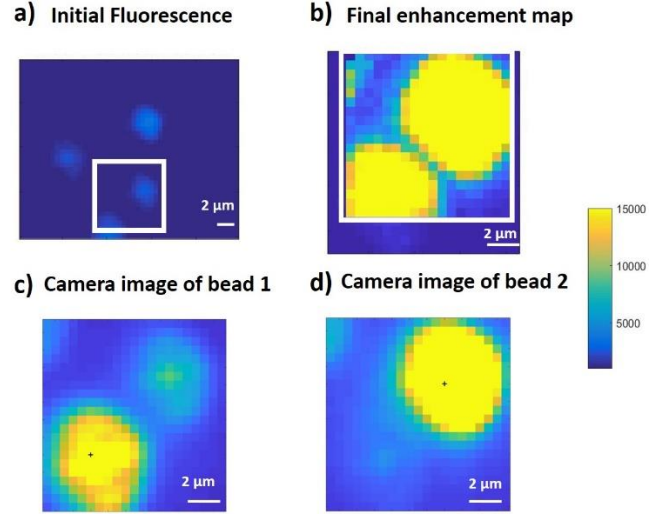

Fig. 3. Enhanced fluorescence signal (a) Initial detected epi-fluorescence in the region of interest where the TM is measured, (b) Enhancement map of the beads after projecting the conjugate transpose of the TM for the two beads within the white square in (a). Focusing process using the TM: (c) camera image (epi-illumination) when the focus is centered on bead 1 (black +), (d) camera image (epi- illumination) when the focus is centered on bead 2 (black +).

According to Ref. [16], the theoretical photoacoustic enhancement one can obtain is equal to $\eta = 0.5 N_{SLM}/N_{modes}$, $N_{modes}$ being the number of speckle grains incident on one absorber. In our experiments, the size of the speckle is close to the size of the beads so we can assume one speckle per bead. We used 64 or 256 active SLM pixels so our theoretical enhancement should be as high as 128. However, in the case of epi-fluorescence detection, the enhancement is further reduced by the back scattering spreading. Still, in experimental conditions, as in our case, one typically obtains about ten times lower enhancements. It should also be pointed out that the reported enhancement values could be underestimated because of possible saturation of the camera. We further note that the enhancement on the speckle field, behind the scatterer, is higher than the enhancement of the epi-fluorescent signal. In a typical epi-fluorescence microscopy experiment, there is no access to monitor the output speckle. In this work, we could place an imaging system behind the sample and obtained an image of the focal spot without removing the beads that shows the focus created is ~2μm at FWHM.

**Discussion and conclusion.** We demonstrated a principle for the measurement of the TM of a scattering medium using epi-fluorescence light as feedback.

The system could be improved by use of a continuous wave laser with lower energy and longer coherence length than the pulsed nanosecond laser. In particular, we noticed some fluctuations in the results due to fluorophore saturation and photobleaching, which affects the assumption that the fluorescence intensity is linear with respect to the illumination light intensity.

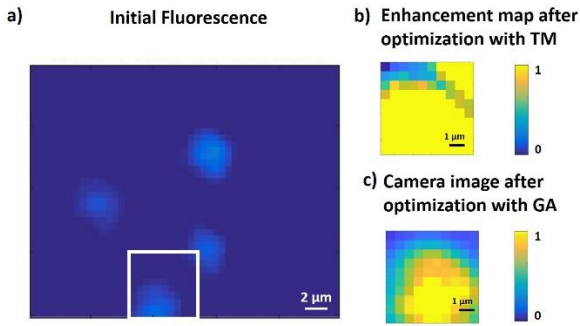

Fig. 4. Comparison of the results obtained with the Transmission Matrix and a Genetic Algorithm. a) Initial epi-fluorescence observed. b) Enhancement map of the beads in the white square after optimization with the TM. c) Camera image of the bead in white square after optimizing with the GA. The enhancement with TM is 15.8, the enhancement with GA is 11.7. We used the same experimental conditions for both algorithms with 256 active pixels on the SLM. The color-bar represents the same scale in all images.

Our method uses an internal reference arm for the phase-shifting interferometric measurement of the TM and is therefore subject to heterogeneities from dark speckles. This could be improved using spiral phases, as in Ref. [11], to increase the performance. Note that in the reflection geometry one cannot use an external reference.

The current limitation of the technique is that it requires weak mixing and sufficient SNR such that the intensity follows a cosine modulation. If the scattering medium is too thick this condition is lost. However, this is a common situation in microscopy through scattering layers. The technique can also work with some extent of mixing of the signals from the various beads. If there is significant overlap among the scattering responses of the various emitters, an additional demixing step should be considered, for instance recording the time signals followed by an independent component analysis [18].

As compared to prior work with photoacoustic TM [16] the use of fluorescence makes the system adaptable to epi-fluorescence microscopy and the reflection geometry could potentially be useful for in-vivo experiments where no optical access is available in transmission. Because the process does not rely on a direct access to the output speckle field, it enables new applications of the TM for microscopy.

We foresee future work to make use of the TM measured with fluorescence light. For example, one could use the measured TM to focus light through the scattering surface and image smaller structures stained with a different dye so one can separate their signal from the brighter beads.

The resolution one can achieve with wavefront shaping depends on the speckle size in the plane of interest. In this work, speckles were about 2 μm at FWHM, which sets the resolution. In order to create a smaller focal spot, we would need a smaller average speckle size. Improving the resolution with epi-fluorescence illumination beyond the size of an individual speckle is a challenge that has yet to be overcome.

**Funding.** We thankfully acknowledge NSF support through awards 1556473 and 1611513.

**Acknowledgment**. We are thankful to Sylvain Gigan (ENS, Paris, France), Thomas Chaigne (Institut Fresnel, Marseille, France), and Simon Labouesse for useful discussions and insights, and to Sakshi Singh for reviewing.